\documentclass[sn-mathphys]{sn-jnl}

\jyear{2021}
\usepackage{amsmath}
\theoremstyle{thmstyleone}
\newtheorem{theorem}{Theorem}

\newtheorem{lemma}{Lemma}

\theoremstyle{thmstyletwo}

\newtheorem{remark}{Remark}
\theoremstyle{thmstylethree}

\begin{document}

\title[{Equivalence Theorem Crossover Design GLM}]{A General Equivalence Theorem for Crossover Designs under Generalized Linear Models}

\author[1]{\fnm{Jeevan} \sur{Jankar}}\email{jeevan.jankar@uga.edu}

\author[2]{\fnm{Jie} \sur{Yang}}\email{jyang06@uic.edu}

\author*[1]{\fnm{Abhyuday} \sur{Mandal}}\email{amandal@stat.uga.edu}

\affil*[1]{\orgdiv{Department of Statistics}, \orgname{University of Georgia}, \city{Athens}, \postcode{30602}, \state{GA}}

\affil[2]{\orgdiv{Department of Mathematics, Statistics, and Computer Science}, \orgname{University of Illinois at Chicago}, \city{Chicago}, \postcode{60607}, \state{IL}}

\newcommand{\bl}[1]{{\color{blue} #1}}
\newcommand{\re}[1]{{\color{red} #1}}
\newcommand{\gr}[1]{{\color{green} #1}}


\abstract{With the help of Generalized Estimating Equations, we identify locally $D$-optimal crossover designs for generalized linear models.  We adopt the variance of parameters of interest as the objective function, which is minimized using constrained optimization to obtain optimal crossover designs. In this case, the traditional general equivalence theorem could not be used directly to check the optimality of obtained designs. In this manuscript, we derive a corresponding general equivalence theorem for crossover designs under generalized linear models.}

\keywords{Approximate Design, Crossover Design, $D$-Optimality, Generalized Estimating Equations, General Equivalence Theorem.}

\maketitle


\section{Introduction}\label{Intro}

Crossover designs, also known as change-over or repeated measurement designs, are widely used in many industrial, medical, and agricultural research. In crossover experiments, the effect of treatments is carried over in the periods following the period of their direct application. Crossover designs have been extensively studied in literature \cite{bib3, bib12, bib14}. Crossover designs have recently been used to address problems outside of medical and agriculture research. In recent years, most corporate offices and organizations have adopted open office spaces over traditional cubical office spaces. \url{Booking. com} conducted an experiment to assess different office spacing efficiency and the case study was first reported in \cite{bib20}. In the absence of literature on optimal cross-over designs for generalized linear models (GLMs), traditional uniform designs are used. Uniform design is optimal under a linear model but they are no longer a good choice for non-Gaussian responses.

Over the years optimal crossover designs for normal responses have been widely studied in the literature, however, there are several examples in real life where responses are not normal and described by GLMs. Recently, \cite{bib13} provided an algorithm to search locally $D$-optimal crossover designs in case of non-normal response, and showed that optimal designs obtained for normal responses can be quite inefficient in the case of GLMs. But, there was no guarantee that the designs obtained by their algorithm were indeed optimal. In this manuscript, we derive a general equivalence theorem specifically for crossover designs under GLMs, which can be used to verify the optimality of proposed designs. Moreover, it provides an alternative that is faster and numerically more stable than the general algorithm proposed in \cite{bib13}.
	
The General Equivalence Theorem is an important tool in optimum experimental designs, which has been widely used for checking for the optimality of designs in terms of the Fisher information matrix \cite{bib2, bib8, bib9, bib10, bib11, bib17, bib28}. Nevertheless, the traditional equivalence theorem does not apply to check the optimality of obtained crossover designs. The optimal crossover designs under GLMs discussed \cite{bib13} are identified using generalized estimating equations (GEEs) and are based on the variance matrix of the parameters of interest. Since the variance matrix is asymptotically connected with the inverse of the Fisher information matrix, it is natural to derive a condition that can be used to check the optimality of designs (see Remark~\ref{Remark_1} for more details).
	
For illustration purposes, we consider two real-life motivating examples. First, we consider an experiment conducted at Booking.com to determine the optimal office design. In the supplementary material, we discuss another motivating example, an experiment conducted to investigate the effects of various dietary starch levels on milk production. \cite{bib15} discussed this dietary example along with the data set used for analysis (for more details see \cite{bib14}).  The design used in both these examples is a $4 \times 4$ Latin Square design with four periods and four treatments.
	
The paper is organized as follows. To set ideas we describe notation and definitions for crossover designs in Section~\ref{Definitions}. In Section~\ref{EqThm_CD} we propose and derive two different versions of the general equivalence theorem for crossover designs. More specifically, in Section~\ref{EqThm_Theta} we use the variance of all parameter estimates as an objective function, and in Section~\ref{EqThm_Tau} we use the variance of treatment effects as an objective function to derive two versions of the theorem. We present illustration in Section~\ref{Illus} and real-life motivating examples in Section~\ref{Motiv_Exmp}.


\section{Notation and Preliminaries}\label{Definitions}

Consider a crossover trial with $t$ treatments, $n$ subjects, and $p$ periods. The responses obtained from these $n$ subjects are denoted as $\boldsymbol{Y_{1}} , \ldots , \boldsymbol{Y_{n}}$, where the response from the $j^{th}$ subject is $ \boldsymbol{Y_{j}} = (Y_{1j} , \ldots , Y_{pj})^\prime $. Let $\mu_{ij}$ denote the mean of a response $Y_{ij}$. To fix ideas, first, consider the following model (see, equation (4.1) in \cite{bib29} and \cite{bib5} for linear model), which models the marginal mean $\mu_{ij}$ for crossover trial as 
\begin{eqnarray}\label{Model}
\textrm{g}(\mu_{ij}) = \eta_{ij} = \lambda + \beta_{i} + \tau_{d(i,j)} + \rho_{d(i-1,j)} ,
\end{eqnarray} 
where $ i = 1,\ldots,p ; j = 1,\ldots,n $; $\lambda$ is the overall mean, $\beta_{i}$ represents the effect of the $i^{th}$ period, $\tau_{s}$ is the direct effect due to treatment $s$ and $\rho_{s}$ is the carryover effect due to treatment $s$, where $s = 1,\ldots,t$ and $g$ is the link function.

In matrix notation, under baseline constraints $\beta_{1} = \tau_{1} = \rho_{1} = 0$  we have $\boldsymbol{\beta} = ( \beta_{2},\ldots,\beta_{p} )^{\prime}$ ,$\boldsymbol{\tau} = ( \tau_{2},\ldots,\tau_{t} )^{\prime}$ and $\boldsymbol{\rho} = ( \rho_{2},\ldots,\rho_{t} )^{\prime}$, which defines the parameter vector $\boldsymbol{\theta} = (\lambda, \beta^{\prime}, \tau^{\prime}, \rho^{\prime})^{\prime}$. The linear predictor corresponding to the $j^{th}$ subject, $\boldsymbol{\eta_{j}} = ( \eta_{1j},\ldots,\eta_{pj} )^{\prime}$, can be written as
\begin{eqnarray*} 
	\boldsymbol{\eta_{j}} &=& \boldsymbol{X_{j}}\boldsymbol{\theta}.
\end{eqnarray*}

The corresponding design matrix $\boldsymbol{X_{j}}$ can be written as $\boldsymbol{X_{j}} = \left[ \boldsymbol{1_{p}}, \boldsymbol{P_{j}}, \boldsymbol{T_{j}}, \boldsymbol{F_{j}} \right ]$, where $\boldsymbol{P_{j}}$ is $p\times(p-1)$ such that $\boldsymbol{P_{j}} = [\boldsymbol{0}_{(p-1)1},\boldsymbol{I}_{p-1}]^\prime$; $\boldsymbol{T_{j}}$ is a $p\times(t-1)$ matrix with its $(i,s-1)^{th}$ entry equal to 1 if subject $j$ receives the direct effect of the treatment $s$ $(\geq 2)$ in the $i^{th}$ period and zero otherwise; $\boldsymbol{F_{j}}$ is a $p\times(t-1)$ matrix with its $(i,s-1)^{th}$ entry equal to 1 if subject $j$ receives the carryover effect of the treatment $s$ $(\geq 2)$ in the $i^{th}$ period and zero otherwise.

If the number of subjects $n$ and the number of periods $p$ are fixed, then the goal is to determine the number of subjects assigned to different treatment sequences through some optimality criterion. As the number of periods $p$ is fixed, each treatment sequence will be of length $p$ and a typical sequence can be written as $\omega = (t_{1},\ldots,t_{p})^{\prime}$ where $t_{i}\in\{1,\ldots,t\}$. Now let $\boldsymbol{\Omega}$ be the set of all such sequences and $n_{\omega}$ denote the number of subjects assigned to sequence $\omega$. Then the total number of subjects $n$ can be written as $n = \Sigma_{\omega\in\boldsymbol{\Omega}}n_{\omega} ,n_{\omega} \geq 0$. A crossover design $\xi$ in approximate theory is specified by the set $\{p_{\omega}, \omega\in\boldsymbol{\Omega}\}$, where $ p_{\omega} = n_{\omega}/n$ is the proportion of subjects assigned to treatment sequence $\omega$. As denoted by Silvey \cite{bib25}, such a crossover design $\xi$ can be written as follows:
\begin{eqnarray}\label{approxdesign}
\xi = \left\{ \begin{array} { l l l l }
{\omega_{1}} & {\omega_{2}} & {\ldots} & {\omega_{k}} \\
{p_{1}} & {p_{2}} & {\ldots} & {p_{k}} \end{array} \right\},
\end{eqnarray}
where $k$ is the number of treatment sequences involved, $\omega_i$ is the $i$th treatment sequence and $p_i$ is the corresponding proportion of units allocated to that support point, such that $\sum_{i = 1} ^ {k} p_{i} = 1, \text { for } i = 1, \ldots, k$. Note \cite{bib13} observed that, in the case of non-uniform allocations, only a few sequences have non-zero proportions. Hence in our illustrations, we consider $\Omega$ to be the collection of only those sequences that have non-zero allocations.

\medskip
Generalized estimating equations are quasi-likelihood equations that allow us to estimate quasi-likelihood estimators \cite{bib21, bib31}. In crossover trials, it is typical to make an assumption that the observations from the same subject are correlated while the observations from different subjects are independent \cite{bib14}. This dependency among repeated observations from the same subject can be modeled by the ``working correlation'' matrix $\boldsymbol{C_\alpha}$, which is a function of correlation coefficient $\alpha.$ If $\boldsymbol{C_\alpha}$ is the true correlation matrix of $\boldsymbol{Y_{j}}$, then from the definition of covariance we can write
\begin{eqnarray*} 
	Cov(\boldsymbol{Y_{j}}) &=& \boldsymbol{D_{j}}^{1/2}\boldsymbol{C_{\alpha}}\boldsymbol{D_{j}}^{1/2},
\end{eqnarray*}
where $\boldsymbol{D_{j}} = diag\Big(Var(Y_{1j}), \ldots, Var(Y_{pj})\Big)$. Let us denote $	Cov(\boldsymbol{Y_{j}})$ by $\boldsymbol{W_{j}}$. In \cite{bib31} (equation (3.1)) it was shown that for repeated measurement models, the generalized estimating equations (GEE) are defined to be 
\begin{eqnarray*} 
	\sum _ { j = 1 } ^ { n } \frac { \partial \boldsymbol{\mu _ { j }} ^ { \prime } } { \partial \boldsymbol{\theta} } \boldsymbol{W _ { j }} ^ { - 1 } \left(\boldsymbol{{ Y } _ { j }} - \boldsymbol{\mu _ { j }} \right) = 0,
\end{eqnarray*}
where $\boldsymbol{\mu _ { j }} = \left( \mu _ { 1 j } , \ldots , \mu _ { p j } \right) ^ { \prime }$ and the asymptotic variance for the GEE estimator $\boldsymbol{\hat{\theta}}$ (see \cite{bib31}, equation (3.2)) is 
\begin{eqnarray}\label{Var_Theta}
{\rm Var} (\boldsymbol{\hat{\theta} }) &=& \left[ \sum _ { j = 1 } ^ { k } np_{j} \frac { \partial \boldsymbol{\mu _ { j }} ^ { \prime } } { \partial \boldsymbol{\theta} } \boldsymbol{W _ { j }} ^ { - 1 } \frac { \partial \boldsymbol{\mu _ { j }} } { \partial \boldsymbol{\theta} } \right] ^ { - 1 } = \boldsymbol{M}^{-1},
\end{eqnarray}
where $\frac { \partial \boldsymbol{\mu _ { j }} ^ { \prime } } { \partial \boldsymbol{\theta} } = \boldsymbol{X_{j}}^{\prime}\text{diag}\left\{(g^{-1})^{\prime}(\eta_{1j}), \ldots, (g^{-1})^{\prime}(\eta_{pj})\right\}$ and $j$ stands for the $j^{th}$ treatment sequence. In Section~\ref{EqThm_CD}, we will define $\boldsymbol{M}$ explicitly for crossover designs. Later, we consider the situation where direct treatment effects are studied specifically. 

\begin{remark}\label{Remark_1}
    Note that the subject effect term is not included in the model (\ref{Model}). In this work, GLM is used to describe the response and hence the Fisher information matrix depends on the model parameters. Since we are considering the local optimality approach, an educated guess for the subject effect, if included, will be needed. But, from a design point of view, the subject effect has to be treated random. In model (\ref{Model}), the link function is used to model only the mean response and hence we are free to choose a variance-covariance matrix. So, instead of including a random subject effect in the model, we choose a working variance-covariance matrix through GEE to capture the effect of a subject (see Appendix A.3 in \cite{bib32}, \cite{bib33}, \cite{bib13} and references therein).
\end{remark}

\begin{remark}\label{Remark_2} The general equivalence theorem describes the optimality criteria in terms of the Fisher information matrix. The information matrix for optimal crossover designs under GLMs is defined as the inverse of the variance-covariance matrix of parameters of interest through GEE, which is easier to obtain and works similarly to the Fisher information matrix. Here we assume that the responses from a particular subject are mutually correlated, while the responses from different subjects are uncorrelated. According to \cite{bib13}, the obtained optimal designs are robust to the choices of such working correlation matrices.
\end{remark}

As mentioned in \cite{bib2}, the general equivalence theorem can be viewed as a consequence of the result that the derivative of a smooth function over an unconstrained region is zero at its minimum. In this manuscript, we derive the general equivalence theorem for crossover designs by calculating the directional derivative of an objective function $\Phi(\xi)$ expressed in terms of $\boldsymbol{M}(\xi)$. Consider $\bar{\xi_{i}}$ to be the design that puts unit mass at the point $x_{i}$, i.e., the design supported only at $x_{i}$, where $i = 1, 2, \ldots, k$. Let $\xi_{i}^{\prime} = (1 - h)\xi + h \bar{\xi_{i}}$. Then the derivative of $\Phi(\xi)$ in the direction $\bar{\xi_{i}}$ or $x_{i}$ in case of $D$-optimal criterion is
$$\phi(x_{i},\xi) = \lim_{h \to 0^{+}} \frac{1}{h}[\Phi(\xi_{i}^{\prime}) - \Phi(\xi_{i})] = - \lim_{h \to 0^{+}} \frac{1}{h}[\ln det(\boldsymbol{M}(\xi_{i}^{\prime})) - \ln det(\boldsymbol{M}(\xi_{i}))],$$
and $\xi$ is $D$-optimal if and only if $min_{i}\phi(x_{i},\xi) = 0$ and $\phi(x_{i},\xi) = 0$ if $p_{\omega_{i}} > 0,$ where this minimum is occurring at the points of support of design.

\medskip
In the case of crossover designs and estimates using generalized estimating equations, a different approach compared to the one mentioned above is needed as the design points are finite and pre-specified for crossover designs. We use the technique used in the supplement materials of \cite{bib30}. Instead of using $\xi_{i}^{\prime} = (1 - h)\xi + h \bar{\xi_{i}} = \xi + h(\bar{\xi_{i}} - \xi)$, they used $\boldsymbol{p_{r}} + u \boldsymbol{\delta_{i}^{(r)}},$ where $\boldsymbol{p_{r}}$ and $\boldsymbol{\delta_{i}^{(r)}}$ are defined below. Therefore, the directional derivative $\phi(u, \boldsymbol{p_{r}})$ of the objective function is equal to $\left.\frac{\partial \Phi(\boldsymbol{p_{r}} + u \boldsymbol{\delta_{i}^{(r)}})}{\partial u} \right \vert_{u = 0}.$ 

\medskip
Here is the outline of the general equivalence theorem in the case of crossover designs. Note that $0 \leq p_{i}< 1$ for $i = 1, \ldots, k$, and since $\sum_{i=1}^{k} p_{i} = 1$ we may assume without any loss of generality that $p_{k} > 0.$  Define $\boldsymbol{p_{r}} = (p_{1}, \ldots, p_{k-1})^{\prime}$, and $\Phi(\boldsymbol{p_{r}}) = -\ln det(\boldsymbol{M}(p_{1}, \ldots, p_{k-1}, 1-\sum_{i=1}^{k-1}p_{i})).$ Let $\boldsymbol{\delta_{i}^{(r)}} = (-p_{1}, \ldots, -p_{i-1}, 1-p_{i}, -p_{i+1}, \ldots, -p_{k-1})^{\prime}$ for $i = 1, \ldots, k-1.$ $\boldsymbol{\delta_{i}^{(r)}}$ are defined in such a way that the determinant $\vert(\boldsymbol{\delta_{1}^{(r)}}, \ldots, \boldsymbol{\delta_{k-1}^{(r)}})\vert = p_{k} \neq 0.$ Hence, $\boldsymbol{\delta_{1}^{(r)}}, \ldots, \boldsymbol{\delta_{k-1}^{(r)}}$ are linearly independent and thus can serve as the new basis of 

$$\boldsymbol{S_{r}} = \{(p_{1}, \ldots, p_{k-1})^{\prime} \vert \sum_{i=1}^{k-1}p_{i} <1, \text{and } p_{i} \geq 0, i = 1, \ldots, k-1\}.$$

\medskip
\noindent Note that negative $\ln det$ is a convex function on a set of positive definite matrices. Hence, $\boldsymbol{p_{r}}$ minimizes $\Phi(\boldsymbol{p_{r}})$ if and only if along each direction $\boldsymbol{\delta_{i}^{r}}$,

\[ \left.\frac{\partial \Phi(\boldsymbol{p_{r}} + u \boldsymbol{\delta_{i}^{(r)}})}{\partial u} \right\vert_{u = 0}
\begin{cases} 
& = 0 \text{ if } p_{i} > 0\\
& \geq 0 \text{ if } p_{i} = 0\\
\end{cases}
\]


\section{Equivalence Theorems for Crossover Designs}\label{EqThm_CD}

As defined earlier, $\boldsymbol{C_\alpha}$ is the ``working correlation'' matrix and hence is a positive definite and symmetric. So, there exists a square matrix $\boldsymbol{R}$ such that $\boldsymbol{C_{\alpha}}^{-1} = \boldsymbol{R}^{T}\boldsymbol{R}$. Then the inverse of the variance of the parameter estimates through GEEs is as follows:

\begin{eqnarray}\label{M}
\boldsymbol{M} &=& \sum _ { j = 1 } ^ { k } np_{j}\frac { \partial \boldsymbol{\mu _ { j }} ^ { \prime } } { \partial \boldsymbol{\theta} } \boldsymbol{W _ { j }} ^ { - 1 } \frac { \partial \boldsymbol{\mu _ { j }} } { \partial \boldsymbol{\theta} } = \sum _ { j = 1 } ^ { k } np_{j}\boldsymbol{X_{j}}^{T}\boldsymbol{G_{j}}\boldsymbol{D_{j}}^{-\frac{1}{2}}\boldsymbol{C_{\alpha}}^{-1}\boldsymbol{D_{j}}^{-\frac{1}{2}}\boldsymbol{G_{j}X_{j}}
\end{eqnarray}
,where $\boldsymbol{G_{j}} = \text{diag}\left\{(g^{-1})^{\prime}(\eta_{1j}), \ldots, (g^{-1})^{\prime}(\eta_{pj})\right\}.$
Equation~\eqref{M} can be further simplified as,

$$\begin{aligned}
\boldsymbol{M} = \sum _ { j = 1 } ^ { k } np_{j}(\boldsymbol{X_{j}}^{*})^{T}(\boldsymbol{X_{j}}^{*}),\\
\end{aligned}$$
where $\boldsymbol{X_{j}}^{*} = \boldsymbol{R}\boldsymbol{D_{j}}^{-\frac{1}{2}}\boldsymbol{G_{j}X_{j}}$.

\subsection{Equivalence Theorem when Objective Function is Variance of Parameter Estimates}\label{EqThm_Theta}

In this section, we present the equivalence theorem for crossover design when the objective function is a determinant of the variance of parameter estimates. We also present a special case of the theorem when there are only two treatment sequences involved in the design.

\begin{theorem}\label{GET_Theta}
	\textbf{(General Equivalence Theorem for Crossover Design when the objective function is $\lvert\boldsymbol{Var(\boldsymbol{\hat{\theta}})}\rvert$):}Consider the design $\xi$ with $k$ treatment sequences as defined in equation~\eqref{approxdesign}. Then,

    \medskip
	(a) The set of optimal designs is convex.
	    
	(b) The design $\xi$ is $D$-optimal if and only if 
	
            	\[ \text{trace}\left(\boldsymbol{X_{i}}^{*}\boldsymbol{M}(\xi)^{-1}\boldsymbol{X_{i}}^{*T}\right)
            	\begin{cases} 
            	& = m \text{ if } p_{i} > 0\\
            	& \leq m \text{ if } p_{i} = 0\\
            	\end{cases},
            	\]
	for each $p_{i} \in [0,1]$, where $p_{i}$ is the allocation corresponding to point $\omega_{i}$ of design $\xi$ for all $i = 1, 2, \ldots, k$, and $m$ is the number of parameters in $\boldsymbol{\theta}.$

\end{theorem}

\noindent Proof of Theorem~\ref{EqThm_Theta}:

\medskip
Let $k$ be the number of treatment sequences involved in the experiment and $\xi$ be any design, then $\Phi(\boldsymbol{M}(\xi)) = -\ln det(\boldsymbol{M}(\xi))$.

\medskip
\noindent Proof of (a):

\medskip
Let $\xi_1^*$ and $\xi_2^*$ be optimal designs i.e.,
    $$\Phi[\boldsymbol{M}(\xi_1^*)] = \Phi[\boldsymbol{M}(\xi_2^*)] = \text{min}_{\xi} \Phi[\boldsymbol{M}(\xi)]$$ 
and let $\xi^* = (1-\gamma)\xi_1^* + \gamma\xi_2^*$, for $0 \leq \gamma \leq 1$. $\Phi[\boldsymbol{M}(\xi)] = -\ln det(\boldsymbol{M}(\xi))$ is convex on set of positive definite matrices \cite{bib6}. Therefore,
    $$\Phi[\boldsymbol{M}(\xi^*)] \leq (1-\gamma)\Phi[\boldsymbol{M}(\xi_1^*)] + \gamma\Phi[\boldsymbol{M}(\xi_2^*)] = \text{min}_{\xi} \Phi[\boldsymbol{M}(\xi)],$$ 
which proves the optimality of $\xi^*$.

\medskip
\noindent Proof of (b):

\medskip
We have $\boldsymbol{p_{r}} = (p_{1}, p_{2}, \ldots, p_{k-1})^{\prime}$ and  $\boldsymbol{\delta_{1}^{(r)}} = (1-p_{1}, -p_{2}, \ldots, -p_{k-1})^{\prime},$

\medskip
\noindent $\boldsymbol{\delta_{2}^{(r)}} = (-p_{1}, 1-p_{2}, \ldots, -p_{k-1})^{'}, \ldots, \boldsymbol{\delta_{k-1}^{(r)}} = (-p_{1}, -p_{2}, \ldots, 1-p_{k-1})^{\prime}.$

\medskip
\noindent Hence, $\boldsymbol{p_{r}} + u \boldsymbol{\delta_{1}^{(r)}} = \left(p_{1} + u (1-p_{1}), (1-u)p_{2}, \ldots, (1-u)p_{k-1}\right)^{\prime},$ 

\noindent $\boldsymbol{p_{r}} + u \boldsymbol{\delta_{2}^{(r)}} = \left((1-u)p_{1}, p_{2} + u (1-p_{2}), \ldots, (1-u)p_{k-1}\right)^{\prime}, \ldots,$

\noindent $\boldsymbol{p_{r}} + u \boldsymbol{\delta_{k-1}^{(r)}} = \left((1-u)p_{1}, (1-u)p_{2}, \ldots, p_{k-1} + u (1-p_{k-1})\right)^{\prime}.$

\medskip
\noindent The determinant of $(\boldsymbol{\delta_{1}^{(r)}}, \cdots, \boldsymbol{\delta_{k-1}^{(r)}})$ is equal to $1 - \left(p_{1} + p_{2} + \cdots + p_{k-1}\right) = p_{k}$.

\noindent Then for design with $k$ treatment sequences we can write $\boldsymbol{M}$ as,
\vspace{-0.3cm}
\begin{multline*}
\boldsymbol{M}(\boldsymbol{p_{r}}) = \sum _ { j = 1 } ^ { k }np_{{j}}(\boldsymbol{X_{j}}^{*})^{T}(\boldsymbol{X_{j}}^{*}) = np_{1}(\boldsymbol{X_{1}}^{*})^{T}(\boldsymbol{X_{1}}^{*}) + np_{2}(\boldsymbol{X_{2}}^{*})^{T}(\boldsymbol{X_{2}}^{*}) + \cdots \\
\hspace{0.8cm} + np_{k-1}(\boldsymbol{X_{k-1}}^{*})^{T}(\boldsymbol{X_{k-1}}^{*}) + n\left(1 - \left(p_{1} + p_{2} + \cdots + p_{k-1}\right)\right)(\boldsymbol{X_{k}}^{*})^{T}(\boldsymbol{X_{k}}^{*})\\
\end{multline*}

\vspace{-0.5cm}
\noindent For illustration purpose consider the direction $\boldsymbol{\delta_{1}^{(r)}}$, and calculations for other directions can be done similarly,
$$\begin{aligned}
\Phi(\boldsymbol{p_{r}} + u \boldsymbol{\delta_{1}^{(r)}}) &= -\ln det\biggl[\boldsymbol{M}\left(\left\{p_{1} + u (1-p_{1}), (1-u)p_{2}, \ldots, (1-u)p_{k-1}\right\}^{\prime}\right)\biggr]\\
\end{aligned}$$
\vspace{-1cm}
\begin{multline*}
= -\ln det\biggl[n\left\{p_{1} + u (1-p_{1})\right\}(\boldsymbol{X_{1}}^{*})^{T}(\boldsymbol{X_{1}}^{*}) \\
\hspace{1cm} + n\left\{(1-u)p_{2}\right\}(\boldsymbol{X_{2}}^{*})^{T}(\boldsymbol{X_{2}}^{*}) + \cdots + n\left\{(1-u)p_{k-1}\right\}(\boldsymbol{X_{k-1}}^{*})^{T}(\boldsymbol{X_{k-1}}^{*}) \\
\hspace{-1.6cm} + n(1-u)\left\{1 - \left(p_{1} + p_{2} + \cdots + p_{k-1}\right)\right\}(\boldsymbol{X_{k}}^{*})^{T}(\boldsymbol{X_{k}}^{*})\biggr]\\
\end{multline*}
\vspace{-1.7cm}
\begin{eqnarray*}
\hspace{-3cm} = -m\ln n - \ln det[\boldsymbol{M}(u, \boldsymbol{p_{r}})] = -m\ln n + \Phi^{(r)}(u),
\end{eqnarray*} where $\boldsymbol{M}(u, \boldsymbol{p_{r}}) = \frac{\boldsymbol{M}(\boldsymbol{p_{r}} + u \boldsymbol{\delta_{1}^{(r)}})}{n},$ and $\Phi^{(r)}(u) = - \ln det[\boldsymbol{M}(u, \boldsymbol{p_{r}})].$

\medskip
\noindent The directional derivative of the above objective function along one specific direction for a design with $k$ treatment sequences can be calculated as follows:
\begin{eqnarray*}
\hspace{-2.5cm} \phi(u,\boldsymbol{p_{r}}) = \frac{\partial \Phi(\boldsymbol{p_{r}} + u \boldsymbol{\delta_{1}^{(r)}})}{\partial u} = \lim_{h \rightarrow 0}\frac{1}{h}\left[\Phi^{(r)}(u + h) - \Phi^{(r)}(u)\right]\\
\end{eqnarray*}
\vspace{-1cm}
\begin{eqnarray*}
\hspace{-3.3cm} = - \lim_{h \rightarrow 0}\frac{1}{h}\biggl\{\ln det\left[\boldsymbol{M}(u+h, \boldsymbol{p_{r}})\right] - \ln det\left[\boldsymbol{M}(u, \boldsymbol{p_{r}})\right]\biggr\}\\
\end{eqnarray*}
\vspace{-1cm}
\begin{multline*}
\hspace{-0.3cm} = - \lim_{h \rightarrow 0}\frac{1}{h}\biggl\{\ln det\biggl[\boldsymbol{M}(u, \boldsymbol{p_{r}}) + h(1-p_{1})\boldsymbol{X_{1}}^{*T}\boldsymbol{X_{1}}^{*} - hp_{2}\boldsymbol{X_{2}}^{*T}\boldsymbol{X_{2}}^{*} - \cdots \\
- hp_{k-1}\boldsymbol{X_{k-1}}^{*T}\boldsymbol{X_{k-1}}^{*} - h\left(1 - \left(p_{1} \cdots + p_{k-1}\right)\right)\boldsymbol{X_{k}}^{*T}\boldsymbol{X_{k}}^{*}\biggr]det\boldsymbol{M}(u, \boldsymbol{p_{r}})^{-1}\biggr\}\\
\end{multline*}
\vspace{-1cm}
$$\begin{aligned}
&= - \lim_{h \rightarrow 0}\frac{1}{h}\biggl\{\ln det\biggl[\boldsymbol{M}(u, \boldsymbol{p_{r}})\boldsymbol{M}(u, \boldsymbol{p_{r}})^{-1} + h\left\{\boldsymbol{X_{1}}^{*T}\boldsymbol{X_{1}}^{*} - \boldsymbol{M}(\boldsymbol{p_{r}})\right\}\boldsymbol{M}(u, \boldsymbol{p_{r}})^{-1}\biggr]\biggr\}\\
&= - \lim_{h \rightarrow 0}\frac{1}{h}\biggl\{\ln det\biggl[\boldsymbol{I}_{p} + h\left\{\boldsymbol{X_{1}}^{*T}\boldsymbol{X_{1}}^{*} - \boldsymbol{M}(\boldsymbol{p_{r}})\right\}\boldsymbol{M}(u, \boldsymbol{p_{r}})^{-1}\biggr]\biggr\}\\
\end{aligned}$$

\noindent Using the approximation of determinant $\text{det}(\boldsymbol{I} + h\boldsymbol{A}) = 1 + h\text{trace}(\boldsymbol{A}) + \mathcal{O}(h^{2})$ \cite{bib4} we get,

\vspace{-0.5cm}
\begin{eqnarray*}
&= - \lim_{h \rightarrow 0}\frac{1}{h}\biggl\{\ln \left(1 + h \text{trace}\left[\left\{\boldsymbol{X_{1}}^{*T}\boldsymbol{X_{1}}^{*} - \boldsymbol{M}(\boldsymbol{p_{r}})\right\}\boldsymbol{M}(u, \boldsymbol{p_{r}})^{-1}\right] + \mathcal{O}(h^{2})\right)\biggr\}\\
\end{eqnarray*}

\vspace{-0.5cm}
\noindent And using $\ln(1+t) = t + \mathcal{O}(t^{2})$ we get,

\vspace{-0.5cm}
$$\begin{aligned}
&= - \lim_{h \rightarrow 0}\frac{1}{h}\biggl\{h\text{trace}\left[(\boldsymbol{X_{1}}^{*T}\boldsymbol{X_{1}}^{*} - \boldsymbol{M}(\boldsymbol{p_{r}}))\boldsymbol{M}(u, \boldsymbol{p_{r}})^{-1}\right] + \mathcal{O}(h^{2})\biggr\}\\
&= - \text{trace}\biggl[(\boldsymbol{X_{1}}^{*T}\boldsymbol{X_{1}}^{*} - \boldsymbol{M}(\boldsymbol{p_{r}}))\boldsymbol{M}(u, \boldsymbol{p_{r}})^{-1}\biggr]\\
&= \text{trace}\left(\boldsymbol{M}(\boldsymbol{p_{r}})\boldsymbol{M}(u, \boldsymbol{p_{r}})^{-1}\right) - \text{trace}\left(\boldsymbol{X_{1}}^{*}\boldsymbol{M}(u, \boldsymbol{p_{r}})^{-1}\boldsymbol{X_{1}}^{*T}\right)\\
\end{aligned}$$

\vspace{-0.5cm}
\begin{eqnarray}\label{EndEqn_Var_Theta}
\left. \frac{\partial \Phi(\boldsymbol{p_{r}} + u \boldsymbol{\delta_{1}^{(r)}})}{\partial u} \right \vert_{u = 0} &= m - \text{trace}\left(\boldsymbol{X_{1}}^{*}\boldsymbol{M}(\boldsymbol{p_{r}})^{-1}\boldsymbol{X_{1}}^{*T}\right)
\end{eqnarray}

\noindent The proof follows by equating the above expression in equation~\eqref{EndEqn_Var_Theta} to zero.

\subsection{Equivalence Theorem when Objective Function is Variance of Treatment Effect Estimates}\label{EqThm_Tau}

As the main interest usually lies in estimating the direct treatment effect contrasts, instead of working with the full variance-covariance matrix of parameters estimate, in this section, we concentrate only on the variance of the estimator of treatment effects ${\rm Var}(\boldsymbol{\hat\tau})$ given as
\begin{eqnarray}\label{Var_Tau}
{\rm Var}(\boldsymbol{\hat\tau}) &=& \boldsymbol{H}{\rm Var}(\boldsymbol{\hat\theta})\boldsymbol{H}^{\prime},
\end{eqnarray}
where $\boldsymbol{H}$ is a $(t-1)\times m$ matrix given by $[\boldsymbol{0}_{(t-1)1},\boldsymbol{0}_{(t-1)(p-1)},\boldsymbol{I}_{t-1},\boldsymbol{0}_{(t-1)(t-1)}]$ and $m=p+2t-2$ is the total number of parameters in $\boldsymbol{\theta}$. Below, we present the equivalence theorem for crossover design when the objective function is a determinant of the variance of treatment effects estimate i.e., the determinant of dispersion matrix.

\begin{lemma}\label{Lemma_Convex}
    Consider function $f: \mathbb{R}_{> 0}^{n}\rightarrow \mathbb{R}_{> 0}$, such that $f(\boldsymbol{x}) = \frac{1}{\prod_{i=1}^{n}x_i}$ where $\boldsymbol{x} = (x_1, x_2, \ldots, x_n)^{\prime} \in \mathbb{R}_{> 0}^{n}$. Then $f(\boldsymbol{x})$ is a strictly convex function.
\end{lemma}

\noindent Proof of Lemma~\ref{Lemma_Convex}:

Let $H$ be the Hessian matrix, i.e., the matrix of second-order partial derivatives. 

\medskip \noindent Then $H = f(x)(D+qq')$, where $D$ is the diagonal matrix with elements $1/(x_1)^2,\ldots, 1/(x_n)^2$ and $q$ is the column vector with elements $1/(x_1),\ldots, 1/(x_n)$. 

\medskip \noindent The lemma follows as $H$ is positive definite. An alternative proof is provided in the supplementary material.

\begin{theorem}\label{GET_Tau}
	\textbf{General Equivalence Theorem for Crossover Design when objective function is $\lvert\boldsymbol{Var(\boldsymbol{\hat{\tau}})}\rvert$: }Consider the design $\xi$ with $k$ treatment sequences as defined in equation~\eqref{approxdesign}. Then,

    \medskip
	(a) The set of optimal designs is convex.
	    
	(b) The design $\xi$ is $D$-optimal if and only if
 
        	\[ \text{trace}\left\{\boldsymbol{A}(\boldsymbol{X_{i}}^{*})^{T}(\boldsymbol{X_{i}}^{*})\right\}
        	\begin{cases} 
        	& = t-1 \text{ if } p_{i} > 0\\
        	& \leq t-1 \text{ if } p_{i} = 0\\
        	\end{cases}
        	\]	
    for each $p_{i} \in [0,1]$, where $\boldsymbol{A} = \boldsymbol{M}^{-1}\boldsymbol{H}^{\prime}\left(\boldsymbol{HM}^{-1}\boldsymbol{H}^{\prime}\right)^{-1}\boldsymbol{HM}^{-1}$, $p_{i}$ is the allocation corresponding to point $\omega_{i}$ of design $\xi$ for all $i = 1, 2, \ldots, k$, and $t$ is number of treatments.
\end{theorem}

\noindent Proof of Theorem~\ref{EqThm_Tau}:

\medskip
Let $k$ be the number of treatment sequences involved in the experiment and $\xi$ be any design, then $\Phi(\boldsymbol{M}(\xi)) = \ln det(\boldsymbol{HM}(\xi)^{-1}\boldsymbol{H}^{\prime})$.

\medskip
\noindent Proof of (a):

\medskip
Let $\xi_1^*$ and $\xi_2^*$ be optimal designs i.e.,
    $$\Phi[\boldsymbol{M}(\xi_1^*)] = \Phi[\boldsymbol{M}(\xi_2^*)] = \text{min}_{\xi} \Phi[\boldsymbol{M}(\xi)]$$ and let $\xi^* = (1-\gamma)\xi_1^* + \gamma\xi_2^*$, for $0 \leq \gamma \leq 1$.
    
\medskip
\noindent Since we are using the $D$-optimality criterion, we prove the following equation~\eqref{eq:HMH_det} to prove the optimality of $\xi^*$.

\vspace{-0.5cm}
    \begin{eqnarray}\label{eq:HMH_det}
    \lvert \boldsymbol{HM}(\xi^*)^{-1}\boldsymbol{H}^{\prime} \rvert \leq (1-\gamma)  \lvert \boldsymbol{HM}(\xi_1^*)^{-1}\boldsymbol{H}^{\prime} \rvert + \gamma  \lvert \boldsymbol{HM}(\xi_2^*)^{-1}\boldsymbol{H}^{\prime} \rvert.
    \end{eqnarray}

\noindent Since both $\boldsymbol{M}(\xi_1^*)$ and $\boldsymbol{M}(\xi_2^*)$ are positive definite, we can find a non-singular matrix $\boldsymbol{O}^{-1}$ such that $\boldsymbol{M}(\xi_1^*) = \boldsymbol{O}\boldsymbol{O}^T$ and $\boldsymbol{M}(\xi_2^*)= \boldsymbol{O}\boldsymbol{\Lambda} \boldsymbol{O}^T$, where $\boldsymbol{\Lambda} = {\rm diag}\{\lambda_1, \ldots, \lambda_m\}$ is a $m\times m$ diagonal matrix (see page 41 \cite{bib22}). In this situation, $\boldsymbol{M}(\xi^*) = \boldsymbol{O}((1-\gamma) \boldsymbol{I}+\gamma\boldsymbol{\Lambda})\boldsymbol{O}^T$. Then \eqref{eq:HMH_det} is equivalent to 
    \begin{equation}\label{eq:GIG_det}
    \lvert \boldsymbol{G}((1-\gamma) \boldsymbol{I}+\gamma\boldsymbol{\Lambda})^{-1}\boldsymbol{G}^T \rvert \leq (1-\gamma)  \lvert \boldsymbol{G}\boldsymbol{G}^T \rvert + \gamma  \lvert \boldsymbol{G}\boldsymbol{\Lambda}^{-1}\boldsymbol{G}^T \rvert,
    \end{equation}
    where $\boldsymbol{G} = \boldsymbol{H}(\boldsymbol{O}^T)^{-1}$. According to Theorem~1.1.2 in \cite{bib9}, 
    \[
    \lvert \boldsymbol{G}((1-\gamma)\boldsymbol{I} + \gamma\boldsymbol{\Lambda})^{-1}\boldsymbol{G}^T \rvert = \sum_{1\leq i_1 < \cdots < i_q \leq m} \lvert \boldsymbol{G}^T[i_1, \ldots, i_q]\rvert ^2 \prod_{l=1}^q \frac{1}{(1-\gamma) + \gamma \lambda_{i_l}},
    \]
    where $\boldsymbol{G}^T[i_1, \ldots, i_q]$ is the $q\times q$ sub-matrix of $\boldsymbol{G}^T$ consisting of the $i_1, \ldots, i_q$ rows of $\boldsymbol{G}^T$. Similarly,
    \[
    (1 - \gamma)  \lvert \boldsymbol{G}\boldsymbol{G}^T \rvert + \gamma  \lvert \boldsymbol{G}\boldsymbol{\Lambda}^{-1}\boldsymbol{G}^T \rvert
    = \sum_{1\leq i_1 < \cdots < i_q\leq m} \lvert \boldsymbol{G}^T[i_1, \ldots, i_q] \rvert^2 \left(1 - \gamma + \gamma  \prod_{l=1}^q  \frac{1}{\lambda_{i_l}}\right).
    \]

\noindent Then \eqref{eq:GIG_det} is true if 
    \begin{equation}\label{eq:questionable}
    \prod_{l=1}^q \frac{1}{(1-\gamma) + \gamma \lambda_{i_l}} \leq 1-\gamma + \gamma  \prod_{l=1}^q  \frac{1}{\lambda_{i_l}}.
    \end{equation}

\noindent Since $f(\boldsymbol{x}) = \frac{1}{\prod_{i=1}^{q}x_i}$ is convex function (from Lemma \ref{Lemma_Convex}), we have $f\left((1 - \gamma)\boldsymbol{1} + \gamma \boldsymbol{\lambda}\right) \leq(1 - \gamma) f(\boldsymbol{1}) + \gamma f(\boldsymbol{\lambda})$, where $\boldsymbol{\lambda} = (\lambda_{i_1}, \cdots, \lambda_{i_q})$ and hence the result follows.

\medskip
\noindent Proof of (b):

\vspace{-0.5cm}
$$\begin{aligned}
\boldsymbol{M}(\boldsymbol{p_{r}}) &= np_{1}(\boldsymbol{X_{1}}^{*})^{T}(\boldsymbol{X_{1}}^{*}) + np_{2}(\boldsymbol{X_{2}}^{*})^{T}(\boldsymbol{X_{2}}^{*}) + \cdots + np_{k-1}(\boldsymbol{X_{k-1}}^{*})^{T}(\boldsymbol{X_{k-1}}^{*}) \\
&+ n\left(1 - \left(p_{1} + p_{2} \cdots + p_{k-1}\right)\right)(\boldsymbol{X_{k}}^{*})^{T}(\boldsymbol{X_{k-1}}^{*}).
\end{aligned}$$

\vspace{-0.5cm}
$$\begin{aligned}
\Phi(\boldsymbol{p_{r}} + u \boldsymbol{\delta_{1}^{(r)}}) &= \Phi\left(\left\{p_{1} + u (1-p_{1}), (1-u)p_{2}, \ldots, (1-u)p_{k-1}\right\}^{\prime}\right)\\
&= \ln det\biggl[\boldsymbol{H}\biggl\{\boldsymbol{M}\left(\left\{p_{1} + u (1-p_{1}), (1-u)p_{2}, \ldots, (1-u)p_{k-1}\right\}^{\prime}\right)\biggr\}^{-1}\boldsymbol{H}^{\prime}\biggr]\\
\end{aligned}$$
\vspace{-0.8cm}
\begin{multline*}
= -(t-1)\ln n + \ln det\biggl[\boldsymbol{H}\biggl\{\left\{p_{1} + u (1-p_{1})\right\}(\boldsymbol{X_{1}}^{*})^{T}(\boldsymbol{X_{1}}^{*}) \\ 
+ \left\{(1-u)p_{2}\right\}(\boldsymbol{X_{2}}^{*})^{T}(\boldsymbol{X_{2}}^{*}) + \cdots + \left\{(1-u)p_{k-1}\right\}(\boldsymbol{X_{k-1}}^{*})^{T}(\boldsymbol{X_{k-1}}^{*}) \\
+ (1-u)\left\{1 - \left(p_{1} + p_{2} + \cdots + p_{k-1}\right)\right\}(\boldsymbol{X_{k}}^{*})^{T}(\boldsymbol{X_{k}}^{*})\biggr\}^{-1}\boldsymbol{H}^{\prime}\biggr]\\
\end{multline*}
\vspace{-1.7cm}
\begin{eqnarray*}
 = -(t-1)\ln n + \ln det\left[\boldsymbol{HM}(u, \boldsymbol{p_{r}})^{-1}\boldsymbol{H}^{\prime}\right] = -(t-1)\ln n + \Phi^{(r)}(u),\\
\end{eqnarray*}

\vspace{-0.7cm}
\noindent where now $\Phi^{(r)}(u) = \ln det\left[\boldsymbol{HM}(u, \boldsymbol{p_{r}})^{-1}\boldsymbol{H}^{\prime}\right]$.

\medskip
\noindent Consider direction $\boldsymbol{\delta_{1}^{(r)}}$, then the directional derivative of the above objective function for a design with $k$ treatment sequences can be calculated as follows:

\vspace{-0.5cm}
\begin{eqnarray*}
\hspace{-2.5cm} \phi(u,\boldsymbol{p_{r}}) &= \frac{\partial \Phi(\boldsymbol{p_{r}} + u\boldsymbol{\delta_{1}^{(r)}})}{\partial u} = \lim_{h \rightarrow 0}\frac{1}{h}\left[\Phi^{(r)}(u + h) - \Phi^{(r)}(u)\right]\\  
\end{eqnarray*}
\vspace{-1.3cm}
\begin{eqnarray*}
= \lim_{h \rightarrow 0}\frac{1}{h}\biggl\{\ln det\left[\boldsymbol{HM}(u+h, \boldsymbol{p_{r}})^{-1}\boldsymbol{H}^{\prime}\right] - \ln det\left[\boldsymbol{HM}(u, \boldsymbol{p_{r}})^{-1}\boldsymbol{H}^{\prime}\right]\biggr\}\\
\end{eqnarray*}
\vspace{-1.3cm}
\begin{multline*}
\hspace{0.03cm} = \lim_{h \rightarrow 0}\frac{1}{h}\biggl\{ \ln det\left[\boldsymbol{H}\left\{(1-\mu-h)\boldsymbol{M}(\boldsymbol{p_{r}}) + (\mu+h)(\boldsymbol{X_{1}}^{*})^{T}(\boldsymbol{X_{1}}^{*})\right\}^{-1}\boldsymbol{H}^{\prime}\right] \\
 - \ln det\left[\boldsymbol{HM}(\mu, \boldsymbol{p_{r}})^{-1}\boldsymbol{H}^{\prime}\right] \biggr\}\\
\end{multline*}
\vspace{-1cm}
\begin{multline*}
\hspace{-0.5cm} = \lim_{h \rightarrow 0}\frac{1}{h}\biggl\{ \ln det\left[\boldsymbol{H}\left\{\boldsymbol{M}(u, \boldsymbol{p_{r}}) - h\left(\boldsymbol{M}(\boldsymbol{p_{r}}) -(\boldsymbol{X_{1}}^{*})^{T}(\boldsymbol{X_{1}}^{*})\right)\right\}^{-1}\boldsymbol{H}^{\prime}\right]\\
 - \ln det\left[\boldsymbol{HM}(u, \boldsymbol{p_{r}})^{-1}\boldsymbol{H}^{\prime}\right] \biggr\}\\
\end{multline*}
\vspace{-1.7cm}
\begin{multline*}
\hspace{-0.5cm} = \lim_{h \rightarrow 0}\frac{1}{h}\biggl\{ \ln det\left[\boldsymbol{H}\left\{\left[\boldsymbol{M}(u, \boldsymbol{p_{r}})\right]\left[\boldsymbol{I} - h\boldsymbol{M}(u, \boldsymbol{p_{r}})^{-1}\left(\boldsymbol{M}(\boldsymbol{p_{r}}) - (\boldsymbol{X_{1}}^{*})^{T}(\boldsymbol{X_{1}}^{*})\right)\right]\right\}^{-1}\boldsymbol{H}^{\prime}\right]\\
\times det\left[\boldsymbol{HM}(u, \boldsymbol{p_{r}})^{-1}\boldsymbol{H}^{\prime}\right]^{-1} \biggr\}\\
\end{multline*}
\vspace{-1.7cm}
\begin{multline*}
\hspace{-0.5cm} = \lim_{h \rightarrow 0}\frac{1}{h}\biggl\{ \ln det\left[\boldsymbol{H}\left\{\left[\boldsymbol{I} - h\boldsymbol{M}(u, \boldsymbol{p_{r}})^{-1}\left(\boldsymbol{M}(\boldsymbol{p_{r}}) -(\boldsymbol{X_{1}}^{*})^{T}(\boldsymbol{X_{1}}^{*})\right)\right]^{-1}\left[\boldsymbol{M}(u, \boldsymbol{p_{r}})\right]^{-1}\right\}\boldsymbol{H}^{\prime}\right]\\
\times det\left[\boldsymbol{HM}(u, \boldsymbol{p_{r}})^{-1}\boldsymbol{H}^{\prime}\right]^{-1} \biggr\}\\
\end{multline*}

\vspace{-0.9cm}
\noindent Assuming $h$ is sufficiently small we use the binomial series expansion $(\boldsymbol{I} + h\boldsymbol{X})^{-1} = \sum_{i=0}^{\infty}(-t\boldsymbol{X})^{i}$ to obtain, 

$$\begin{aligned}
\phi(u,\boldsymbol{p_{r}}) &= \lim_{h \rightarrow 0}\frac{1}{h}\left\{\ln det\left[\boldsymbol{I} + h\boldsymbol{B} + \mathcal{O}(h^{2})\right]\right\},\\
\end{aligned}$$

\medskip
\noindent $\boldsymbol{B} = \boldsymbol{HM}(u, \boldsymbol{p_{r}})^{-1}\left[\boldsymbol{M}(\boldsymbol{p_{r}}) -(\boldsymbol{X_{1}}^{*})^{T}(\boldsymbol{X_{1}}^{*})\right]\boldsymbol{M}(u, \boldsymbol{p_{r}})^{-1}\boldsymbol{H}^{\prime}\left[\boldsymbol{HM}(u, \boldsymbol{p_{r}})^{-1}\boldsymbol{H}^{\prime}\right]^{-1}.$

\bigskip
\noindent Using $\ln det\left[\boldsymbol{I} + h\boldsymbol{B} + \mathcal{O}(h^{2})\right] = h\text{trace}(\boldsymbol{B}) + \mathcal{O}(h^{2})$ \cite{bib26},

\vspace{-0.5cm}
\begin{multline*}
\phi(u,\boldsymbol{p_{r}}) = \text{trace}\biggl\{\boldsymbol{HM}(u, \boldsymbol{p_{r}})^{-1}\left[\boldsymbol{M}(\boldsymbol{p_{r}}) -(\boldsymbol{X_{1}}^{*})^{T}(\boldsymbol{X_{1}}^{*})\right]\boldsymbol{M}(u, \boldsymbol{p_{r}})^{-1}\boldsymbol{H}^{\prime}\\
\times\left[\boldsymbol{HM}(u, \boldsymbol{p_{r}})^{-1}\boldsymbol{H}^{\prime}\right]^{-1}\biggr\}\\
\end{multline*}
\vspace{-1.7cm}
\begin{multline*}
\phi(u,\boldsymbol{p_{r}}) \vert_{u = 0} = \text{trace}\biggl\{\boldsymbol{HM}(\boldsymbol{p_{r}})^{-1}\left[\boldsymbol{M}(\boldsymbol{p_{r}}) -(\boldsymbol{X_{1}}^{*})^{T}(\boldsymbol{X_{1}}^{*})\right]\boldsymbol{M}(\boldsymbol{p_{r}})^{-1}\boldsymbol{H}^{\prime}\\
\times \left[\boldsymbol{HM}(\boldsymbol{p_{r}})^{-1}\boldsymbol{H}^{\prime}\right]^{-1}\biggr\}\\
\end{multline*}
\vspace{-1.7cm}
\begin{multline*}
\hspace{1.8cm} = \text{trace}\biggl\{\boldsymbol{I}_{(t-1)} - \boldsymbol{HM}(\boldsymbol{p_{r}})^{-1}(\boldsymbol{X_{1}}^{*})^{T}(\boldsymbol{X_{1}}^{*})\boldsymbol{M}(\boldsymbol{p_{r}})^{-1}\boldsymbol{H}^{\prime}\\
\times\left[\boldsymbol{HM}(\boldsymbol{p_{r}})^{-1}\boldsymbol{H}^{\prime}\right]^{-1}\biggr\}\\
\end{multline*}
\vspace{-1.3cm}
$$\begin{aligned}
&= (t-1) - \text{trace}\biggl\{\boldsymbol{HM}(\boldsymbol{p_{r}})^{-1}(\boldsymbol{X_{1}}^{*})^{T}(\boldsymbol{X_{1}}^{*})\boldsymbol{M}(\boldsymbol{p_{r}})^{-1}\boldsymbol{H}^{\prime}\left(\boldsymbol{HM}(\boldsymbol{p_{r}})^{-1}\boldsymbol{H}^{\prime}\right)^{-1}\biggr\}\\
\end{aligned}$$
\vspace{-0.5cm}
\begin{eqnarray}\label{EndEqn_Var_Tau}
\hspace{-1.1cm}	= (t-1) - \text{trace}\biggl\{\left[\boldsymbol{M}^{-1}\boldsymbol{H}^{\prime}\left(\boldsymbol{H}\boldsymbol{M}^{-1}\boldsymbol{H}^{\prime}\right)^{-1}\boldsymbol{H}\boldsymbol{M}^{-1}\right](\boldsymbol{X_{1}}^{*})^{T}(\boldsymbol{X_{1}}^{*})\biggr\}
\end{eqnarray}

\noindent The proof follows by equating the above expression in equation~\eqref{EndEqn_Var_Tau} to zero.

\subsection{Illustration}\label{Illus}	

To illustrate the results of the above general equivalence theorems, we consider a design space $\{AB, BA\}$ has $k = 2, p = 2$. Since we are considering a local optimality approach, for illustration purposes we assume that the parameter values are $\boldsymbol{\theta} = (\lambda, \beta_{2}, \tau_{B}, \rho_{B})^{\prime} =$ $(0.5, -1.0, 4.0, -2.0)^{\prime}.$ Note that we need to assume parameter values before calculating the optimal proportions. Considering the AR(1) correlation structure with $\alpha = 0.1$, i.e., 
$$\boldsymbol{C_\alpha} = \Big(\alpha^{\vert i - i^{\prime} \vert}\Big) = \left(\begin{array}{rr} 1 & \alpha \\
\alpha & 1 \\ \end{array}\right),$$
for the assumed parameter values the optimal proportions are $p_1 = p_2 = 0.5$.

\begin{center}
	\begin{figure}[h]
		\centering 
		\begin{tabular}{cc}
			\includegraphics[scale=0.43]{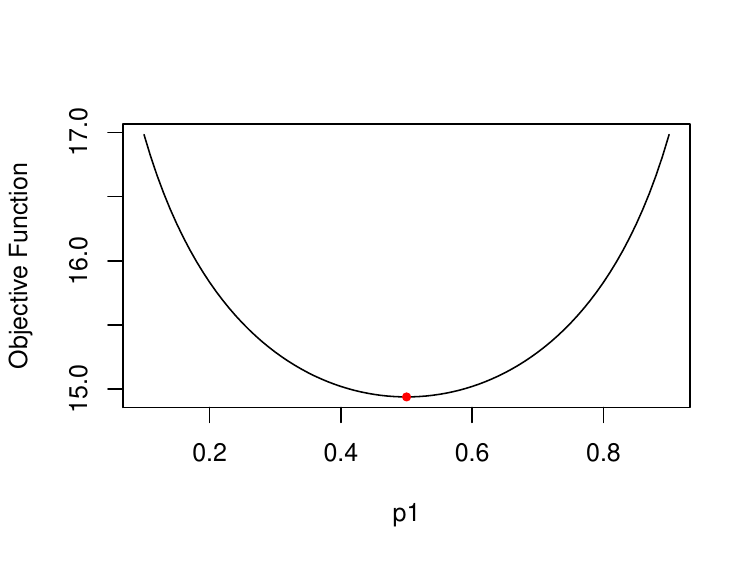} & \includegraphics[scale=0.43]{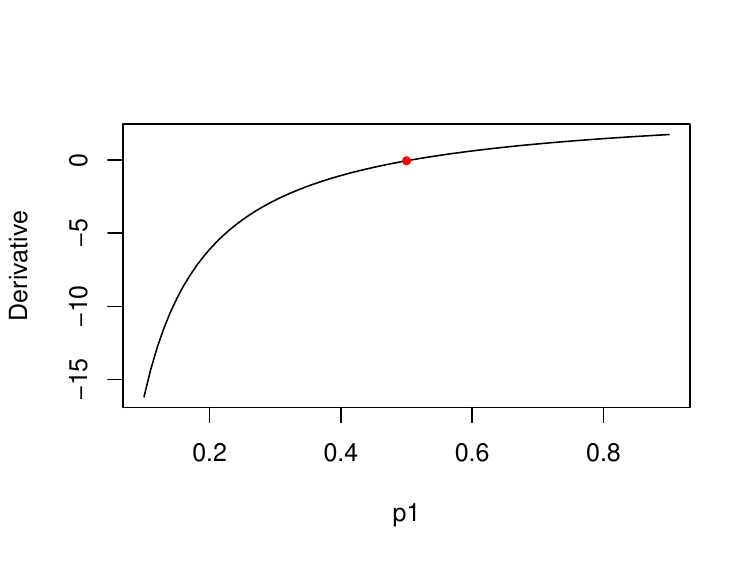}\\
		\end{tabular}
		\caption{Objective function and its directional derivative for designs with two treatment sequences.}
		\label{Fig:ObjFn_Theta_k=2}
	\end{figure}
\end{center}

\medskip
The graph of the objective function, $\Phi(p_{1}) = -\ln det(\boldsymbol{M}(p_{1}))$ and its directional derivative $\textit{trace}\left(\boldsymbol{X_{1}}^{*}\boldsymbol{M}(p_{1})^{-1}\boldsymbol{X_{1}}^{*T}\right) - m$ w.r.t $p_{1} \in [0,1]$ are shown in Figure \ref{Fig:ObjFn_Theta_k=2}.

\vspace{-0.3cm}
Graphs in Figure \ref{Fig:ObjFn_Theta_k=2} verify that the minimum of the objective function is located at $p_{1} = 0.5$ and directional derivative is zero at $p_{1} = 0.5.$ Using Theorem~\ref{GET_Theta}, we conclude that for assumed values of parameters, design

\[
\xi = \left\{ \begin{array} { c c }
AB & BA\\
{ 0.5 } & { 0.5 } \end{array} \right\}
\] 
is the $D$-optimal design when the objective function is $Var(\boldsymbol{\hat{\theta}})$.

\begin{center}
	\begin{figure}[h]
		\centering 
		\begin{tabular}{cc}
			\includegraphics[scale=0.45]{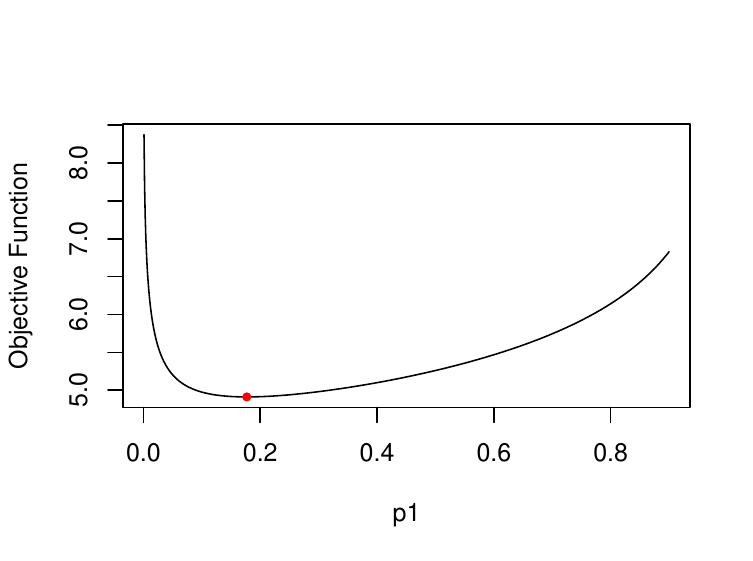} & \includegraphics[scale=0.45]{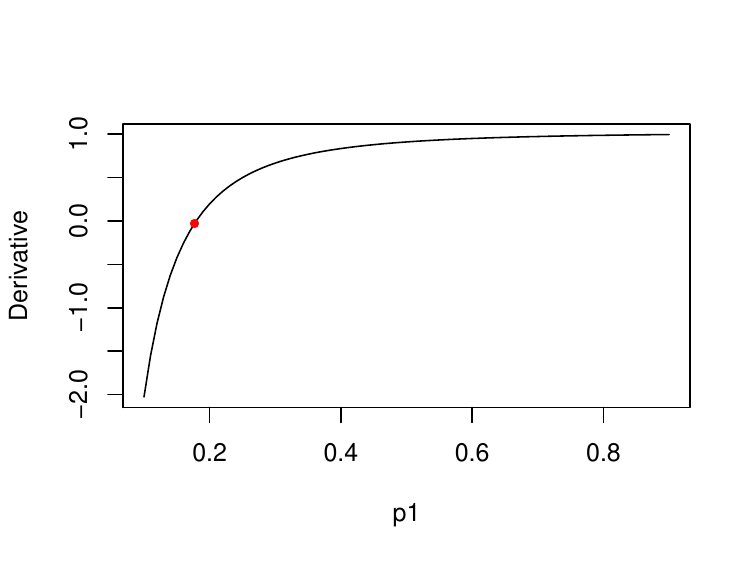}\\
		\end{tabular}
		\caption{Objective function and its directional derivative for designs with two treatment sequences.}
		\label{Fig:ObjFn_Tau_k=2}
	\end{figure}
\end{center}

\medskip
Considering $Var(\boldsymbol{\hat{\tau}})$ as the objective function, the graph of the objective function, $\Phi(p_{1}) = \ln det[\boldsymbol{HM}(p_{1})^{-1}\boldsymbol{H}^{'}]$ and it's directional derivative $\textit{trace}\left\{\boldsymbol{A}(\boldsymbol{X_{1}}^{*})^{T}(\boldsymbol{X_{1}}^{*})\right\}  - (t-1)$ w.r.t $p_{1} \in [0,1]$ are shown in Figure \ref{Fig:ObjFn_Tau_k=2}.

Graphs in Figure \ref{Fig:ObjFn_Tau_k=2} verify that the minimum of the objective function is located at $p_{1} = 0.177$ and directional derivative is zero at $p_{1} = 0.177.$ Using Theorem~\ref{GET_Tau}, we conclude that for assumed values of parameters, design 

\[
\xi = \left\{ \begin{array} { c c }
AB & BA\\
{ 0.177 } & { 0.823 } \end{array} \right\}
\] 
is the $D$-optimal design when the objective function is $Var(\boldsymbol{\hat{\tau}})$.


\section{Real Example}\label{Motiv_Exmp}

In this section, we look at the application of the above equivalence theorems to a real-life example discussed earlier. We obtain the $D$-optimal designs by solving a system of equations given by the general equivalence theorems when the objective functions are $Var(\boldsymbol{\hat{\theta}})$ and $Var(\boldsymbol{\hat{\tau}})$, respectively.	

\subsection{Work Environment Experiment}
Consider the data obtained from the work environment experiment conducted at Booking.com \cite{bib20}. In recent years, many corporate offices and organizations have adopted open office spaces over traditional cubical office spaces. Since there were no previous studies to examine the effects of office designs in workspaces, Booking.com conducted an experiment to assess different office spacing efficiency.

\begin{center}
	\begin{figure}[h]
		\centering 
		\begin{tabular}{cc}
			\includegraphics[scale=.25]{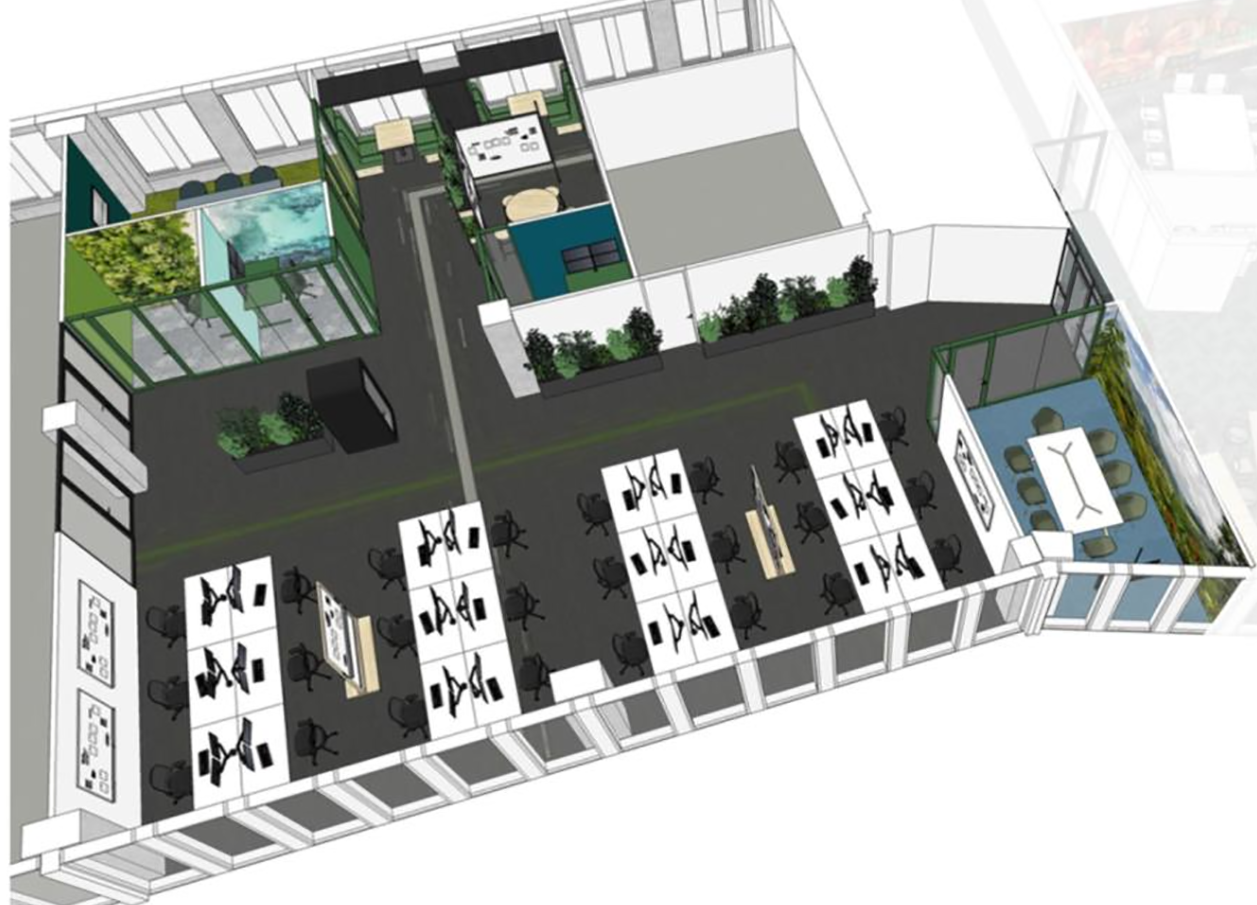} & \includegraphics[scale=.2]{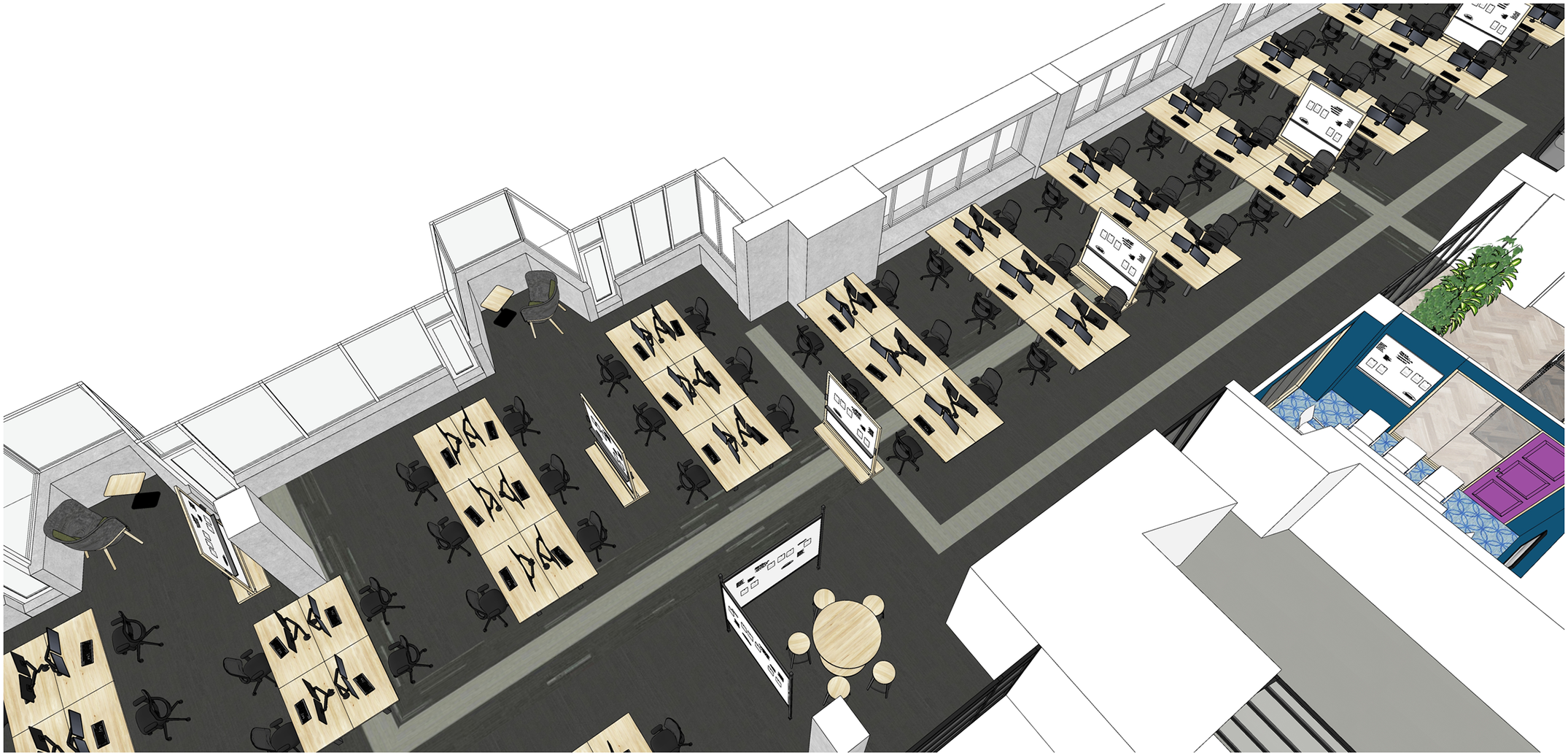}\\
			$A -$ ACT & $B -$ OPEN \\
			& \\
			\includegraphics[scale=.2]{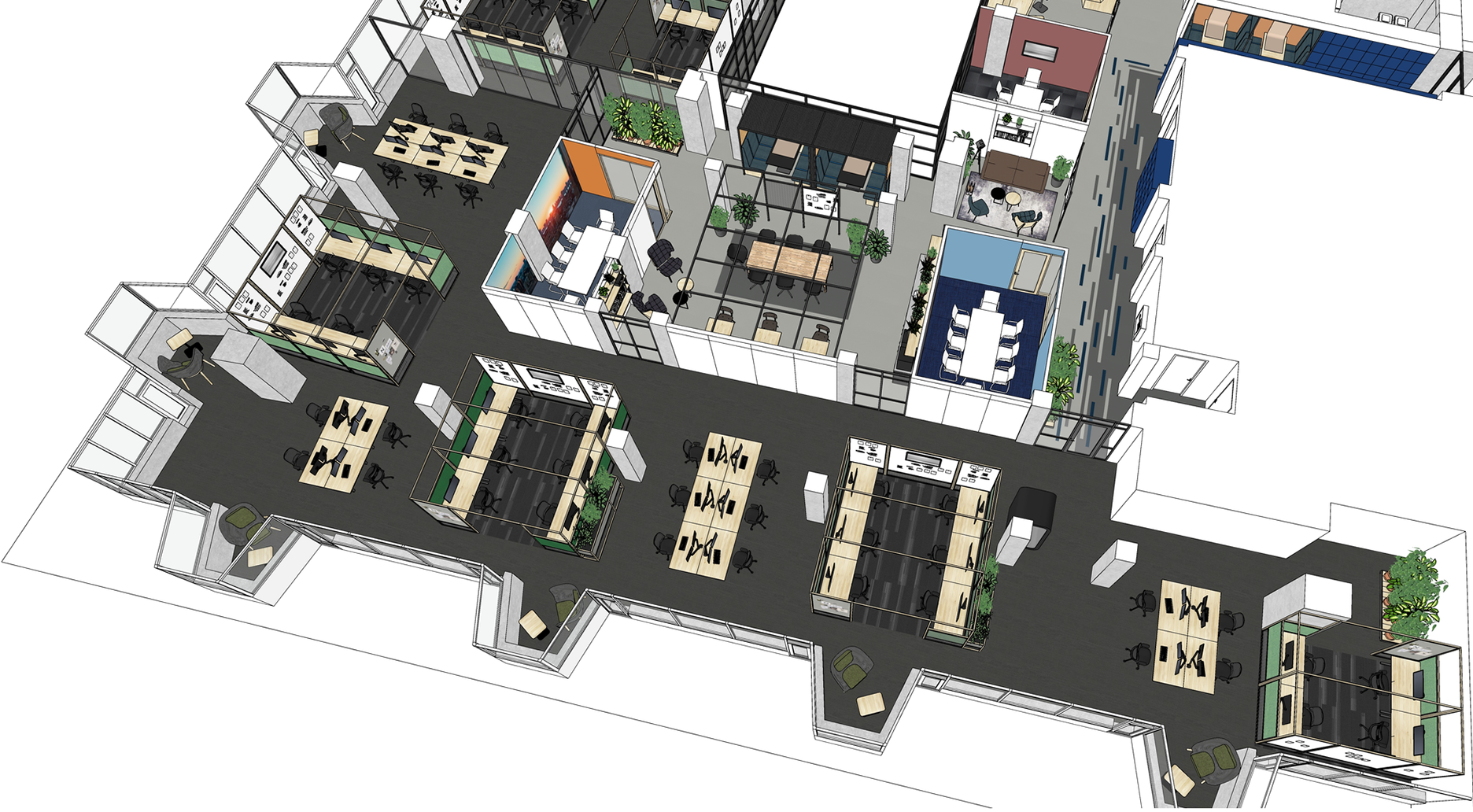} & \includegraphics[scale=.25]{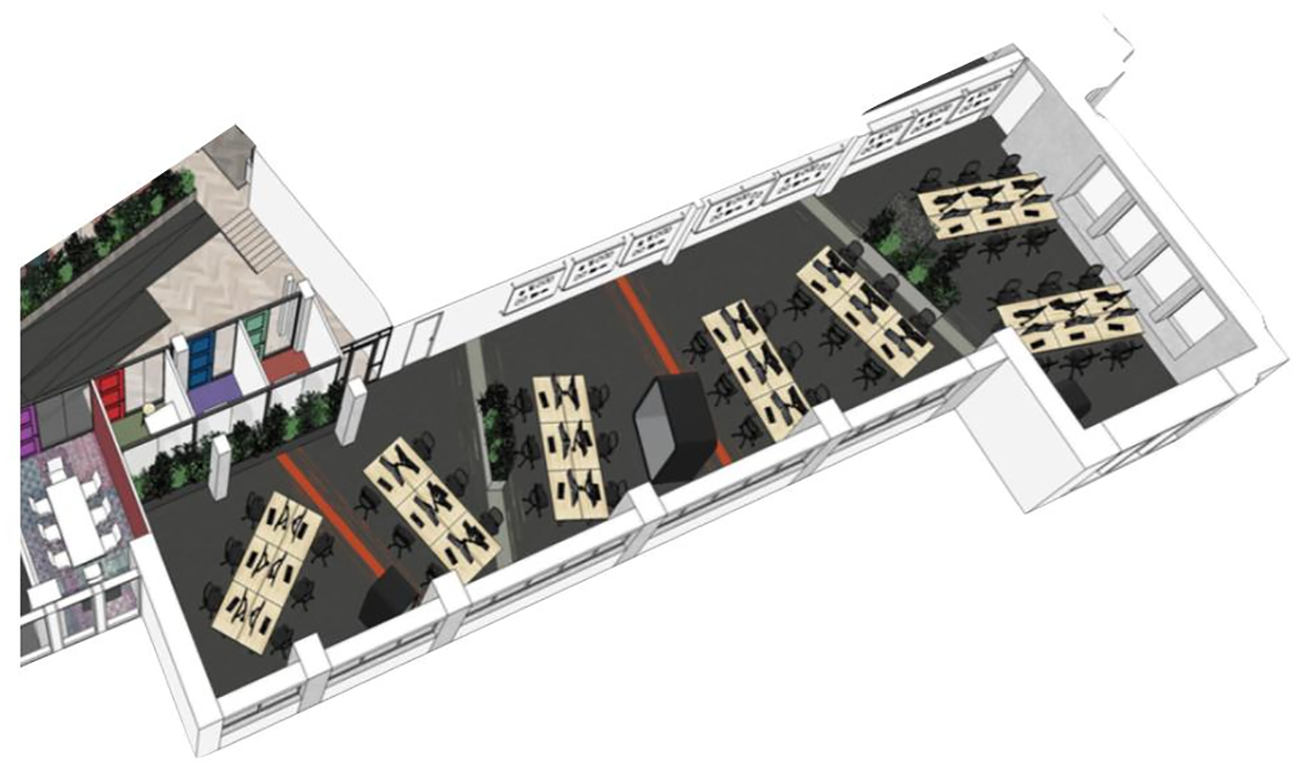} \\
			$C -$ TEAM & $D -$ ZONE. \\
		\end{tabular}
		\caption{The four office designs involved in Work Environment Experiment}
		\label{Fig:Office_Design}
	\end{figure}
\end{center}

In this experiment, there were a total of $n = 288$ participants. Participants were divided into four groups $G_1, G_2, G_3, G_4$ with each group having an equal number of 72 individual participants. It is essentially a uniform crossover design with $p = 4$ periods and $t = 4$ treatments. Periods were named Wave 1, Wave 2, Wave 3, and Wave 4, where each Wave had a duration of 2 weeks. The four treatments involved in this experiment are office designs named $A$ (Activity-Based), $B$ (Open Plan), $C$ (Team Offices), and $D$ (Zoned Open Plan), as shown in the Figure~\ref{Fig:Office_Design}. During the experiment, each group is exposed to different treatments over different periods depending on its treatment sequence. At a particular given period, there was no interaction between subjects from different groups. A Latin square design (see, for example, \cite{bib29}) of order four has been used to determine the sequence of exposure so that no group was exposed to the conditions in the same order as any other group. The design is shown below in Table~\ref{LTD}. A total of $m = 23$ covariates was involved in the experiment, but we consider only the most important ones in our fitted model. 

The images are reproduced from the manuscript \cite{bib20}, under Creative Commons Attribution license (\url{https://creativecommons.org/licenses/by/4.0/}). 
\begin{table}[h]\caption{Latin square design}\label{LTD}
	{\scriptsize	
		\begin{center}
			\begin{tabular}{ |c|cccc|}
				\hline 
				&&&&\\[-7pt]
				Groups $\Rightarrow$ & $G_1$ & $G_2$ & $G_3$ & $G_4$ \\
				Period $\Downarrow$ & $(BADC)$ & $(CDAB)$ & $(DBCA)$ & $(ACBD)$ \\
				\hline
				&&&&\\[-7pt]
				Wave 1 & OPEN & TEAM & ZONE & ACT  \\
				Wave 2 & ACT  & ZONE & OPEN & TEAM \\
				Wave 3 & ZONE & ACT  & TEAM & OPEN \\
				Wave 4 & TEAM & OPEN & ACT  & ZONE \\
				\hline 
			\end{tabular}
		\end{center}
	}
\end{table}

For illustration purposes, consider the response $commit$ $count$ to illustrate the optimal crossover design for the Poisson response. The commit count is the number of commits submitted to the main git repository by each subject. In the fitted model, we examine three primary predictors: $area$, $wave$, and $carryover$. Here, $area$ represents the direct treatment effect, $wave$ denotes the period effect, and $carryover$ represents the effect of the treatment from the previous period. To illustrate the local optimality approach, we assume specific parameter values $\boldsymbol{\theta} = $ ($2.0$, $0.3$, $0.8$, $-0.1$, $-2.0$, $0.40$, $-2.0$, $-1.0$, $0.3$, $-1.0$)$^\prime$, which lead to non-uniform allocations using the log link function and AR(1) correlation structure with $\alpha = 0.1$. 

According to Theorem~\ref{EqThm_Theta}, the $D$-optimal design, i.e., the optimal proportions, can be obtained by solving the following system of equations instead of performing constrained optimization:

$$\text{trace}\left(\boldsymbol{X_{i}}^{*}\boldsymbol{M}(\boldsymbol{p_{r}})^{-1}\boldsymbol{X_{i}}^{*T}\right) = 10,$$ 
for $i = 1, 2, 3, 4.$ The resulting $D$-optimal design is the same as the one obtained through constrained optimization, indicating that the design is given by:

\[
\xi = \left\{ \begin{array} { l l l l }
{BADC} & {CDAB} & {DBCA} & {ACBD}\\
{0.2375} & {0.2894} & {0.2246} & {0.2485} \end{array} \right\}
\] 
is the $D$-optimal design when the objective function is $Var(\boldsymbol{\hat{\theta}})$.

\medskip
Similarly, according to Theorem~\ref{EqThm_Tau}, for the objective function $Var(\boldsymbol{\hat{\tau}})$, the $D$-optimal design can be obtained by solving the following system of equations:

\vspace{-0.5cm}
$$\text{trace}\left\{\left[\boldsymbol{M}(\boldsymbol{p_{r}})^{-1}\boldsymbol{H}^{\prime}\left(\boldsymbol{HM}(\boldsymbol{p_{r}})^{-1}\boldsymbol{H}^{'}\right)^{-1}\boldsymbol{HM}(\boldsymbol{p_{r}})^{-1}\right](\boldsymbol{X_{i}}^{*})^{T}(\boldsymbol{X_{i}}^{*})\right\} = 3,$$
for $i = 1, 2, 3, 4.$ Again, the resulting $D$-optimal design is the same as the one obtained through constrained optimization, indicating that the design is given by:
\[
\xi = \left\{ \begin{array} { l l l l }
{BADC} & {CDAB} & {DBCA} & {ACBD}\\
{0.2900} & {0.2963} & {0.1734} & {0.2403} \end{array} \right\}
\]
is the $D$-optimal design.

\begin{remark}
In  \cite{bib13}, we study the effect of misspecification of working correlation structures on optimal design. We calculate optimal designs under two choices of unknown parameters for a misspecified working correlation structure. Then we calculate relative $D$-efficiency under two parameter choices.  The relative $D$-efficiency under two parameter choices suggests that the effect of variance misspecification on the local optimal designs is minimal. We also study the performance of proposed locally optimal designs via sensitivity study in terms of relative loss of efficiency for choosing assumed parameter values instead of true parameter values. The relative loss of efficiency increases as we move away from true parameter values. However, \textbf{Fig 6.} in \cite{bib13} suggest that this loss of efficiency does not go beyond 2$\%$. We also calculate the optimal designs with all 24 sequences, by considering AR (1) correlation structure and different values of $\alpha$. We observe that in the case of non-uniform allocations, the optimal design has more zeros than non-zero proportions; and these allocations do not vary a lot as $\alpha$ changes, particularly for the sequences where we have zero allocations.
\end{remark}


\section{Summary and Conclusion}\label{Concl}
In many experiments in real life, uniform designs are typically used. Uniform designs are optimal in the case of a linear model i.e., when the response is normally distributed. However, in situations where responses are non-normal, the obtained optimal proportions are not necessarily uniform. In this manuscript, we derive an expression for the general equivalence theorem to check for the optimality of identified locally $D$-optimal crossover designs for generalized linear models. The equivalence theorem provides us with a system of equations that can calculate optimal proportions without performing constrained optimization of the objective function. We derive two different versions of the general equivalence theorem, one with the objective function $Var(\boldsymbol{\hat{\theta}})$ and the other with the objective function $Var(\boldsymbol{\hat{\tau}})$. We illustrate the application of these equivalence theorems on two real-life examples and obtain the same set of optimal proportions by solving the system of equations as obtained by performing constrained optimization. In our future work, we plan to use the Bayesian approach to avoid guessing the values of unknown parameters.

\bigskip\noindent {\it Funding:} This research is in part supported by NSF Grant DMS-2311186.

\bigskip\noindent {\it Conflict of Interests:} There is no conflict of interest.


\bibliography{sn-bibliography}

\begin{thebibliography}{9}

\bibitem{bib1} K. Afsarinejad. Repeated measurement designs - a review. Communications in Statistics - Theory and Methods., 19(11):3985–4028, 1990.

\bibitem{bib2} Anthony Atkinson, Alexander Donev, and Randall Tobias. Optimum Experimental Designs, with SAS. 01 2007.

\bibitem{bib3} R. Bailey and J. Kunert. On optimal crossover designs when carryover effects are proportional to direct effects. Biometrika, 93:613–625, 09 2006.

\bibitem{bib4} Folkmar Bornemann. On the numerical evaluation of Fredholm determinants. Mathematics of Computation, 79(270):871–915, 2010.

\bibitem{bib5} Mausumi Bose and Aloke Dey. Optimal Crossover Designs. World Scientific, 2009.

\bibitem{bib6} Stephen P Boyd and Lieven Vandenberghe. Convex Optimization. Cambridge University Press, 2004.

\bibitem{bib7} K. C. Carriere and Rong Huang. Crossover designs for two-treatment clinical trials. Journal of Statistical Planning and Inference, (87):125–134, 2002.

\bibitem{bib8} V.V. Fedorov. The design of experiments in the multiresponse case. Theory Probab. Appl., (16):323–332, 1971.

\bibitem{bib9} V.V. Fedorov. Theory of Optimal Experiment. Academic Press, New York, 1972.

\bibitem{bib10} V.V. Fedorov and S.L. Leonov. Optimal Design for Nonlinear Response Models. Chapman \& Hall/CRC, 2014.

\bibitem{bib11} V.V. Fedorov and M.B. Malyutov. Optimal designs in regression problems. Math. Operat. Statist., (3):281–308, 1972.

\bibitem{bib12} A. S. Hedayat and Min Yang. Optimal and efficient crossover designs for comparing test treatments with a control treatment. The Annals of Statistics, 33(2):915 – 943, 2005.

\bibitem{bib13} Jeevan Jankar, Abhyuday Mandal, and Jie Yang. Optimal crossover designs for generalized linear models. Journal of Statistics Theory and Practice, 14(23), 2020.

\bibitem{bib14} B. Jones and M.G. Kenward. Design and Analysis of Cross-Over Trials. Chapman \& Hall/CRC Monographs on Statistics \& Applied Probability. Taylor \& Francis, 2014.

\bibitem{bib15} M. G. Kenward and B. Jones. Alternative approaches to the analysis of binary and categorical repeated measurements. Journal of Biopharmaceutical Statistics, 2(2):137–170, 1992. PMID: 1300210.

\bibitem{bib16} Ronald P. Kershner and Walter T. Federer. Two-treatment crossover designs for estimating a variety of effects. Journal of the American Statistical Association, 76(375):612–619, 1981.

\bibitem{bib17} J. Kiefer and J. Wolfowitz. The equivalence of two extremum problems. Canadian Journal of Mathematics, 12:363–366, 1960.

\bibitem{bib18} Eugene M. Laska and Morris Meisner. A variational approach to optimal two-treatment crossover designs: Application to carryover-effect models. Journal of the American Statistical Association, 80(391):704–710, 1985.

\bibitem{bib19} J. N. S. Matthews. Recent developments in crossover designs. International Statistical Review / Revue Internationale de Statistique, 56(2):117–127, 1988.

\bibitem{bib20} J Pitchforth, E Nelson White, M van den Helder, and Oosting W. The work environment pilot: An experiment to determine the optimal office design for a technology company. PLoS ONE, 15(5), 2020.

\bibitem{bib21} Ross L. Prentice. Correlated binary regression with covariates specific to each binary observation. Biometrics, 44(4):1033–1048, 1988.

\bibitem{bib22}C. Radhakrishna Rao. Linear Statistical Inference and its Applications. John
Wiley Sons, Ltd, 1973.

\bibitem{bib23} Karlin S and Studden W.J. Tchebycheff systems: With applications in analysis and statistics. New York: Interscience Publishers, 1966.

\bibitem{bib24} S.S. Senn. Cross-over Trials in Clinical Research. Statistics in Practice. Wiley, 2002.

\bibitem{bib25} S D Silvey. Optimal Design. 1980.

\bibitem{bib26} Christopher S.Withers and Saralees Nadarajah. Expansion for functions of determinants of power series. Canadian Applied Mathematics Quarterly, 18(1):107–114, 2010.

\bibitem{bib27} Cini Varghese, AR Rao, and V. Sharma. Robustness of Williams Square changeover designs. Metrika, 55:199–208, 06 2002.

\bibitem{bib28} P. Whittle and M.B. Malyutov. Some general points in the theory of optimal experimental design. J. Roy. Statist. Soc. Ser., (35):123–130, 1973.

\bibitem{bib29} C.F.J. Wu and M. Hamada. Experiments: Planning, Analysis, and Parameter Design Optimization. Wiley India Pvt. Limited, 2009.

\bibitem{bib30} Jie Yang, Abhyuday Mandal, and Dibyen Majumdar. Optimal design for $2^k$ factorial experiments with binary response. Statistica Sinica, 26(1):385–411, 2016.

\bibitem{bib31} S. Zeger, K. Liang, and P. Albert. Models for longitudinal data: a generalized estimating equation approach. Biometrics, 44 4:1049–60, 1988.

\bibitem{bib32} Christopher Eager and Joseph Roy. Mixed Effects Models are Sometimes Terrible, 2017.

\end{thebibliography}

\clearpage
\newpage

\begin{center}
    { {\Large A General Equivalence Theorem for Crossover \\ \medskip Designs under Generalized Linear Models}} 

     \bigskip

    Jeevan Jankar, Jie Yang and Abhyuday Mandal
\end{center}

\section*{Supplementary Material}

\subsubsection*{Alternative Proof of Lemma~\ref{Lemma_Convex}:}

We use the first-order condition of a convex function stated in equation (3.2) of \cite{bib6}, \textit{A differentiable function $f$ defined on a convex domain is convex if and only if $f(\boldsymbol{x}) \geq f(\boldsymbol{y}) + \nabla f(\boldsymbol{y})^{T}(\boldsymbol{x}-\boldsymbol{y})$ hold for all $\boldsymbol{x}, \boldsymbol{y}$ in the domain.}

\bigskip
Let $\boldsymbol{x} = (x_1, x_2, \ldots, x_n)^{\prime}$, $\boldsymbol{y} = (y_1, y_2, \ldots, y_n)^{\prime} \in \mathbb{R}_{> 0}^{n}$.

\bigskip
Then, we want to show
$$
f(\boldsymbol{x}) \geq f(\boldsymbol{y}) + \nabla f(\boldsymbol{y})^{T}(\boldsymbol{x}-\boldsymbol{y}),
$$
\begin{eqnarray*}
  \text{i.e., to show, } && \frac{1}{\prod_{i=1}^{n}x_i} - \frac{1}{\prod_{i=1}^{n}y_i} \geq \left[
\begin{array}{ccc}
\frac{-1}{y_{1}^2 y_2 \dots y_n}  \ldots  \frac{-1}{y_{1} \dots y_{n-1} y_{n}^2}
\end{array}
\right] \left[
\begin{array}{c}
x_1 - y_1 \\ x_2 - y_2 \\ \dots \\ x_n - y_n
\end{array}
\right],    \\\\
\text{i.e., to show, } && \frac{1}{\prod_{i=1}^{n}x_i} - \frac{1}{\prod_{i=1}^{n}y_i} \geq 
\frac{y_1 - x_1}{y_{1}^2 y_2 \dots y_n} + \dots + \frac{y_n - x_n}{y_{1} \dots y_{n-1} y_{n}^2},  \\\\
\text{i.e., to show, } && \frac{1}{\prod_{i=1}^{n}x_i} - \frac{1}{\prod_{i=1}^{n}y_i} \geq \frac{n}{\prod_{i=1}^{n}y_i} - \sum_{i=1}^{n} \frac{x_i}{y_{1} \dots y_{i}^2 \dots y_n}, \\\\
\text{i.e., to show, } && {\frac{1}{\prod_{i=1}^{n}x_i} + \sum_{i=1}^{n} \frac{x_i}{y_{1} \dots y_{i}^2 \dots y_n}} \geq \frac{n+1}{\prod_{i=1}^{n}y_i}.
\end{eqnarray*}

\medskip
Since $\boldsymbol{x}, \boldsymbol{y} > 0$, the LHS is the mean of $(n+1)$ positive terms. By applying AM $\geq$ GM inequality, the result follows.

\bigskip

\subsection*{Real-Life Example: Dairy Dietary Experiment}
In the introduction section of the manuscript, we provided a brief overview of a dairy dietary experiment that aimed to investigate the impact of various dietary starch levels on milk production. The experiment followed a four-period four-treatment trial design, as first proposed by \cite{bib15}. To administer the order in which diets were received by cows, a Latin square design with four treatment sequences was employed.

In this specific example, the researchers obtained binary responses from a four-period crossover trial. They allocated four treatment sequences to a group of eighty different subjects across the four periods. At the end of each period, the efficacy measurement of each subject was recorded as success or failure, depending on whether a diet was effective or not, resulting in a joint outcome at the end of the trial. The dataset contained the four pre-determined treatment sequences $\Omega = {ABCD, BDAC, CADB, DCBA}$ along with the joint outcomes of the four different periods for each subject following a specific treatment sequence. The Latin square design used in the above experiment is an example of $k = 4, p = 4$. To illustrate the local optimality approach, we assume specific parameter values $\boldsymbol{\theta} = $ ($-2$, $0.25$, $0$, $0.75$, $1$, $5$, $-1.5$, $-3.5$, $2.75$, $0.75$)$^{\prime}$, which lead to non-uniform allocations with the logit link function and AR(1) correlation structure with $\alpha = 0.1$. 

According to Theorem~\ref{EqThm_Theta}, the $D$-optimal design, i.e., optimal proportions, can be obtained by solving the following system of equations instead of performing constrained optimization:

$$\text{trace}\left(\boldsymbol{X_{i}}^{*}\boldsymbol{M}(\boldsymbol{p_{r}})^{-1}\boldsymbol{X_{i}}^{*T}\right) = 10,$$ 
for $i = 1, 2, 3, 4.$ The resulting $D$-optimal design is the same as the one obtained through constrained optimization, indicating that the design given by:

\[
\xi = \left\{ \begin{array} { l l l l }
{ABCD} & {BDAC} & {CADB} & {DCBA}\\
{0.3540} & {0.2108} & {0.2726} & {0.1626} \end{array} \right\}
\] 
is the $D$-optimal design when objective function is $Var(\boldsymbol{\hat{\theta}})$.

Similarly, according to Theorem~\ref{EqThm_Tau}, for the objective function $Var(\boldsymbol{\hat{\tau}})$, the $D$-optimal design can be obtained by solving the following system of equations:

$$\text{trace}\left\{\left[\boldsymbol{M}(\boldsymbol{p_{r}})^{-1}\boldsymbol{H}^{\prime}\left(\boldsymbol{HM}(\boldsymbol{p_{r}})^{-1}\boldsymbol{H}^{'}\right)^{-1}\boldsymbol{HM}(\boldsymbol{p_{r}})^{-1}\right](\boldsymbol{X_{i}}^{*})^{T}(\boldsymbol{X_{i}}^{*})\right\} = 3,$$
for $i = 1, 2, 3, 4.$ Again, the obtained $D$-optimal design is the same as the one obtained through constrained optimization, indicating that the design given by

\[
\xi = \left\{ \begin{array} { l l l l }
{ABCD} & {BDAC} & {CADB} & {DCBA}\\
{0.1725} & {0.2482} & {0.2225} & {0.3586} \end{array} \right\}
\] 
is the $D$-optimal design.

\end{document}